\documentclass[sigconf]{acmart}
\acmConference[EASE 2026]{The 30th International Conference on Evaluation and Assessment in Software Engineering}{Tue 9 - Fri 12 June 2026}{Glasgow, United Kingdom}
\usepackage{graphicx} 
\usepackage{tabularx}
\usepackage{booktabs}
\usepackage{caption}
\usepackage{graphicx}
\acmBooktitle{Proceedings of the 30th International Conference on Evaluation and Assessment in Software Engineering (EASE), 2026, June 9--12, 2026, Glasgow, United Kingdom}

\begin{document}
\title{Applications of Causality in Software Testing: A Rapid Review}
\author{Tiancheng Ma}
    \affiliation{%
        \institution{University of Tennessee}
        \city{Knoxville}
        \state{TN}
        \country{USA}
}
\email{tma10@vols.utk.edu}

\author{Nasir U. Eisty}
\affiliation{%
	   \institution{University of Tennessee}
	   \city{Knoxville}
	   \state{TN}
	   \country{USA}
}
\email{neisty@utk.edu}
\date{June 2026}

\begin{abstract}
Causal inference offers a principled framework for understanding how interventions influence software behavior, yet its adoption in software testing remains fragmented across different tasks and research communities. In this rapid review, we systematically analyze 27 studies that apply causal reasoning to software testing activities such as debugging, fairness assessment, and performance evaluation. We organize the literature using a layered causal inference pipeline, spanning causal representation, structure discovery, identification, and effect estimation, to reveal how existing work maps onto fundamental causal reasoning stages. Our analysis shows a concentration of research on identification and estimation, while representation and discovery techniques are underexplored in testing contexts. We also identify cross-layer challenges, including model misspecification, untested assumptions, and limited empirical evaluation, which hinder practical application. Based on these insights, we propose a research agenda that highlights underrepresented opportunities for advancing causal methods in software testing. This structured perspective aims to unify disparate contributions and guide future empirical and methodological work in the intersection of causality and software testing.





\end{abstract}


\maketitle

\section{Introduction}
As software systems become increasingly complex by incorporating interacting components, adaptive behaviors, and data-driven elements, testing faces persistent challenges. These include the absence of reliable functional or safety oracles~\cite{10132179, monjezi2023informationtheoretictestingdebuggingfairness, 10.1145/3607184}, intricate and interacting input--output behaviors~\cite{10132179, foster2025usingcausalinferencetest, 10.1145/3696630.3731613}, and difficulty interpreting outcomes under confounding effects and spurious correlations~\cite{10.1109/ICSE-NIER58687.2023.00018, 10.1145/1831708.1831717, 10.1109/ASE56229.2023.00106}. Practical constraints, such as output uncertainty~\cite{10.1145/3607184}, limited observability~\cite{foster2025usingcausalinferencetest}, and restricted controllability of feasible and repeatable interventions~\cite{10.1109/ASE56229.2023.00106}, further complicate reliable and reproducible testing.

Many of these challenges stem from a fundamental difficulty: distinguishing genuine causal relationships from mere statistical associations. Traditional testing and debugging techniques often rely on correlational signals extracted from execution traces or test outcomes, which can be distorted by hidden confounders or interacting program components. As a result, observed associations may not reflect true causal influence, limiting the interpretability and reliability of testing conclusions.

Causal inference provides a principled framework for reasoning about cause--effect relationships rather than correlations~\cite{10.1109/ICSE-NIER58687.2023.00018, 10.1145/3607184, 10.1145/1831708.1831717}. Instead of asking whether two variables are statistically associated, it asks whether and how an intervention on one variable would change another~\cite{10.1145/3635709, monjezi2023informationtheoretictestingdebuggingfairness, 10.1109/ASE56229.2023.00106}. Within the Structural Causal Model (SCM) framework, causal assumptions are encoded using directed graphs and structural equations, enabling formal reasoning about interventions, confounding control, and counterfactual queries~\cite{10132179, 10.1109/ICSE-NIER58687.2023.00018}. These capabilities align closely with core testing needs, particularly in result interpretation, root cause analysis, fairness assessment, and performance debugging.

Recent studies have begun integrating causal reasoning into software testing workflows, applying techniques such as causal modeling of program entities, intervention-based analysis (e.g., input perturbation or fault injection), and causal effect estimation. However, this body of work remains fragmented across application domains and testing phases. Existing contributions differ in how they represent causal structures, operationalize interventions, and estimate effects, and they rely on heterogeneous assumptions whose implications are not yet systematically analyzed. As a result, it remains unclear how causal inference is positioned within testing, which parts of the causal reasoning pipeline are well studied, and where methodological and empirical gaps persist.

To address this gap, we conduct a rapid review of primary studies at the intersection of causal inference and software testing. Synthesizing 27 studies, we analyze (i) how causal inference has been positioned within testing workflows, (ii) what techniques and assumptions have been adopted, (iii) what cross-layer risks and limitations constrain practical application, and (iv) what research opportunities emerge from current patterns.





This paper makes the following contributions:

\begin{itemize}
    \item \textbf{Systematic synthesis.} We present the first structured rapid review of 27 primary studies at the intersection of causal inference and software testing.

    \item \textbf{Layered taxonomy.} We propose a four-layer causal inference perspective: representation, discovery, identification, and estimation, and relate it to testing phases.

    \item \textbf{Risk analysis.} We identify recurring cross-layer methodological risks affecting validity and scalability.

    \item \textbf{Research agenda.} We outline key directions for advancing causal methods in software testing.
\end{itemize}

\section{Goal and Research Questions}
\label{goal}
The goal of this study is to characterize the current state of research at the intersection of causal inference and software testing through a rapid review, and to articulate a forward-looking perspective for advancing causal methods in testing. Specifically, we aim to systematically identify the techniques that have been proposed, the types of systems and contexts in which they have been evaluated, and the methodological and empirical gaps that remain. 

Therefore, we formulate the following research questions:

\begin{description}
    \item[RQ1:] How is causal inference positioned in software testing?
    \item[RQ2:] What causal inference techniques and underlying assumptions are adopted in software testing?
    \item[RQ3:] What challenges limit the application of causal inference in software testing?
    \item[RQ4:] What are the research gaps and future directions?

\end{description}

\section{Search Strategy}
\begin{figure*}[t]
    \centering
    \includegraphics[width=\textwidth]{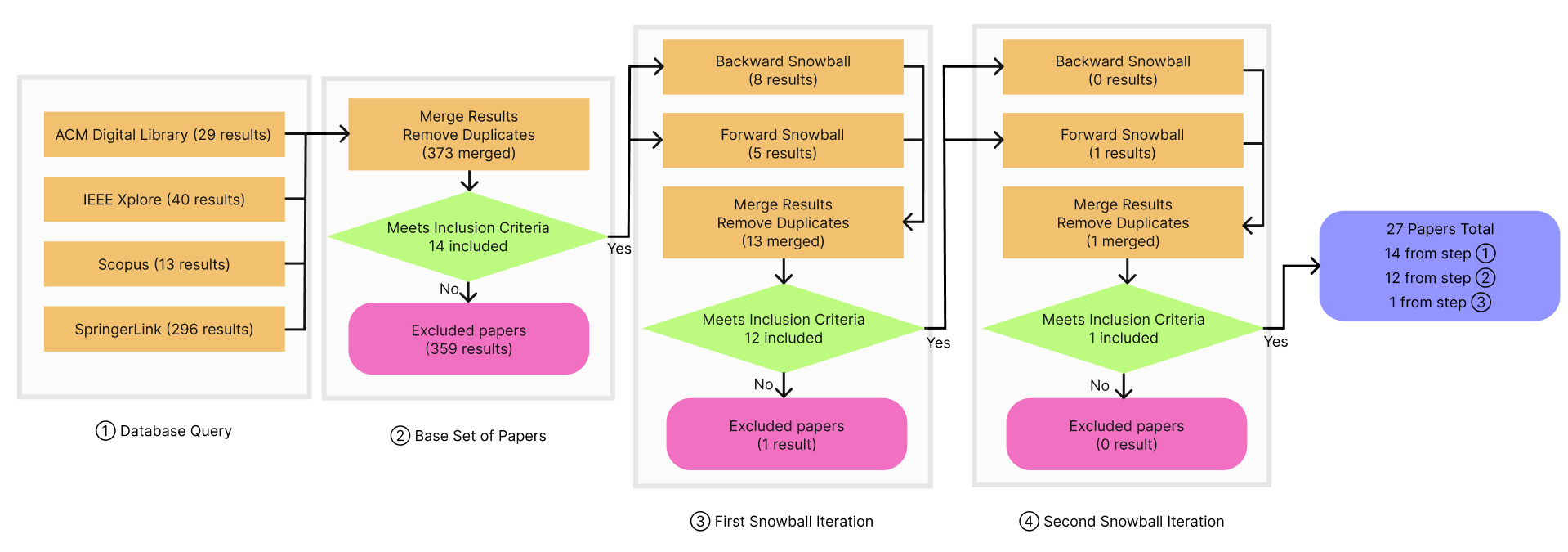}
    \caption{Results from the initial database query \textcircled{1} were filtered by our inclusion and exclusion criteria \textcircled{2}. Two rounds of forward and backward snowballing were completed \textcircled{3}, \textcircled{4} to yield a final set of studies.}
    \label{fig:pipeline}
\end{figure*}

We conducted a rapid review following the guidelines of Cartaxo et al.~\cite{cartaxoRoleRapidReviews2018} and complemented the search process with backward and forward snowballing as recommended by Wohlin~\cite{wohlinGuidelinesSnowballingSystematic2014}. The objective was to systematically synthesize existing research on the application of causal inference in software testing.

Searches were performed in the \textit{ACM Digital Library}, \textit{IEEE Xplore}, \textit{Scopus}, and \textit{SpringerLink}. We included peer-reviewed studies published between 2010 and 2025 that explicitly addressed the use of causal inference techniques within software testing contexts and answered at least one of our research questions. The initial database queries identified 373 unique papers. After applying the inclusion and exclusion criteria through title, abstract, and full-text screening, 14 studies were retained. Backward and forward snowballing yielded an additional 13 relevant papers, resulting in a final dataset of 27 primary studies (see Figure~\ref{fig:pipeline}).

The final search strings are as follows:

\begin{description}
    \item[ACM] (
  (Title:("software testing" OR "software verification" OR "software validation")
   OR
   Abstract:("software testing" OR "software verification" OR "software validation"))
  AND
  (Title:(causality OR "causal inference" OR "causal graph")
   OR
   Abstract:(causality OR "causal inference" OR "causal graph"))
)
    \item[IEEE Xplore] ("software testing" OR "software verification" OR "software validation")
AND
(causality OR "causal inference" OR "causal graph") 

    \item[Scopus] TITLE-ABS(
  ("software testing" OR "software verification" OR "software validation")
  AND
  ("causality" OR "causal inference" OR "causal graph")
)
    \item[Springer Link] (   (title:("software testing" OR "software verification" OR "software validation")    OR    abstract:("software testing" OR "software verification" OR "software validation"))   AND   (title:("causality" OR "causal inference" OR "causal graph")    OR    abstract:("causality" OR "causal inference" OR "causal graph")) )

\end{description}

\subsection{Selection Procedure}

The first author performed each step detailed in Figure~\ref{fig:pipeline} while the second author confirmed the results. Any disagreements were resolved through discussion. In steps \textcircled{2}, \textcircled{3}, and \textcircled{4}, papers were screened in two rounds. The first round screened each potential paper based on its title and abstract. The second round consisted of a full-text review to select the most relevant studies.

\subsection{Extraction Procedure}

Data relevant to our research questions were extracted from each paper and grouped by research question by the first author for analysis and synthesis. The second author reviewed the results to ensure all relevant data were correctly identified.

\section{Results}
In this section, we report the findings of the rapid review and systematically address the research questions introduced in Section~\ref{goal}.

\subsection{RQ1: How is causal inference positioned in software testing?}

To understand how causal inference is integrated into software testing, we synthesized the included studies along two complementary dimensions: (i) the \emph{functional role} played by causal reasoning within a testing approach, and (ii) the \emph{testing phase} in which it is primarily applied. Table~\ref{tab:rq1_positioning} summarizes the distribution of studies across these dimensions.

\subsubsection{Functional Role}

Our analysis reveals three recurring functional roles of causal inference in software testing.

\begin{itemize}
    \item \textbf{Enhancement mechanism.} In the majority of studies, causal inference is introduced to strengthen an existing testing or debugging technique without fundamentally altering its paradigm. Here, causal reasoning augments traditional statistical or heuristic methods, for example, by improving fault localization accuracy or reducing confounding bias in debugging metrics.
    \item \textbf{Testing methodological foundation.} In a smaller but distinct set of studies, causal inference serves as the theoretical backbone of the testing approach. In these cases, the overall framework is explicitly grounded in causal modeling principles, and testing activities, such as adequacy assessment or performance analysis, are structured around formal procedures for estimating interventions and effects.
    \item \textbf{Testing paradigm reformulation.} A limited number of studies go further by reframing software testing itself as a causal inference problem. Rather than treating causal reasoning as an auxiliary tool, these approaches conceptualize testing as the systematic estimation of causal effects, shifting the focus from observational association to intervention-based reasoning.
\end{itemize}

\subsubsection{Testing Phase}

To further contextualize the integration of causal inference, we mapped studies to testing phases defined by Ammann and Offutt~\cite{10.5555/3133461}. 

\begin{itemize}
    \item \textbf{Test design.} Causal inference supports the construction of causally informative test cases by identifying relevant variables, specifying intervention targets, and selecting configurations that enable meaningful effect estimation.
    \item \textbf{Test execution.} Some approaches embed causal reasoning during execution by collecting runtime data suitable for causal analysis and performing controlled interventions, such as input perturbations or fault injections, to observe treatment–outcome relationships.
    \item \textbf{Result interpretation.} Result interpretation is the most common integration point. Here, causal inference is used to distinguish genuine causal relationships from spurious correlations, control for confounding variables, and quantify causal effects observed during testing.
    \item \textbf{Debugging.} Causal inference is positioned as a debugging aid, enabling root cause analysis, isolation of faulty components, and explanation of system behavior through counterfactual reasoning and causal effect estimation.
\end{itemize}

Overall, causal inference is most commonly applied downstream in interpretation and debugging, while fewer studies integrate it into earlier testing phases such as design or execution.

\begin{table*}[t]
\centering
\small
\caption{Positioning of Causal Inference in Software Testing}
\label{tab:rq1_positioning}
\begin{tabularx}{0.92\textwidth}{|l|X|}
\hline
\textbf{Category} & \textbf{Elements and Supporting Studies} \\ \hline

Role
& \textbf{Enhancement mechanism:} 
\cite{10.1145/1831708.1831717}
\cite{10.1145/2025113.2025136}
\cite{7102597}
\cite{10.1145/3696630.3731613}
\cite{11260638}
\cite{10.1145/3318464.3389694}
\cite{Feyzi_2019}
\cite{foster2025usingcausalinferencetest}
\cite{10638595}
\cite{10.1145/3635709}
\cite{10.1109/ICSE-NIER58687.2023.00018}
\cite{6227169}
\cite{10.1109/ASE56229.2023.00106}
\cite{10.1145/3377811.3380377}
\cite{kucuk2021improvingfaultlocalizationintegrating}
\cite{lee2021causalprogramdependenceanalysis}
\cite{Li2025EnhancingFL}
\cite{10.1145/3663529.3663807}
\cite{monjezi2023informationtheoretictestingdebuggingfairness}
\cite{oh2021effectivelysamplinghigherorder}
\cite{9240635}
\cite{6569724}\\

& \textbf{Testing paradigm reformulation:} 
\cite{10556461}
\cite{Poskitt_2023}
\\ 

& \textbf{Testing framework or methodology foundation:} 
\cite{10.1145/3607184}
\cite{10.1145/3696630.3731613}
\cite{10638595}
\cite{10.1145/3635709}
\cite{6606573}
\\ 
\hline

Test phase
& \textbf{Test design:}
\cite{10.1145/3607184}
\cite{10132179}
\cite{foster2025usingcausalinferencetest}
\cite{10556461}
\cite{10.1145/3635709}
\cite{10.1109/ICSE-NIER58687.2023.00018}
\cite{oh2021effectivelysamplinghigherorder}
\cite{Poskitt_2023}\\

& \textbf{Test execution:}
\cite{10.1145/3635709}
\cite{lee2021causalprogramdependenceanalysis}
\\

& \textbf{Result interpretation:}
\cite{10.1145/1831708.1831717}
\cite{10.1145/2025113.2025136}
\cite{7102597}
\cite{10.1145/3607184}
\cite{10.1145/3696630.3731613}
\cite{Feyzi_2019} 
\cite{foster2025usingcausalinferencetest}
\cite{10638595}
\cite{10.1145/3635709}
\cite{10.1109/ASE56229.2023.00106}
\cite{10.1145/3377811.3380377}
\cite{kucuk2021improvingfaultlocalizationintegrating}
\cite{lee2021causalprogramdependenceanalysis}
\cite{monjezi2023informationtheoretictestingdebuggingfairness}
\cite{9240635}
\cite{6569724}
\cite{6606573}\\

& \textbf{Debugging:}
\cite{10.1145/1831708.1831717}
\cite{10.1145/2025113.2025136}
\cite{7102597}
\cite{11260638}
\cite{10.1145/3318464.3389694}
\cite{Feyzi_2019}
\cite{6227169}
\cite{10.1109/ASE56229.2023.00106}
\cite{10.1145/3377811.3380377}
\cite{kucuk2021improvingfaultlocalizationintegrating}
\cite{lee2021causalprogramdependenceanalysis}
\cite{Li2025EnhancingFL}
\cite{10.1145/3663529.3663807}
\cite{monjezi2023informationtheoretictestingdebuggingfairness}
\cite{9240635}
\cite{6569724}
\cite{6606573}
\\ \hline
\end{tabularx}
\end{table*}

\subsection{RQ2: What causal inference techniques and underlying assumptions are adopted in software testing?}

\subsubsection{\textbf{Causal Inference Techniques}}
Based on the synthesis of the primary studies, we derived a layered taxonomy of causal inference techniques in software testing.
Grounded in Pearl’s SCM framework~\cite{10.5555/1642718}, we organize these techniques into four stages of the causal inference workflow: \textit{causal representation}, \textit{causal structure discovery}, \textit{causal identification}, and \textit{causal effect estimation}.

\textbf{\textit{1. Causal Representation Layer.}}
The causal representation layer specifies how dependencies among program entities are encoded as causal structures. Representative formalisms include Structural Causal Models (SCMs) \cite{10.1145/3635709}, Directed Acyclic Graphs (DAGs) \cite{10.1145/3607184}, and program dependence graphs (PDGs) interpreted as causal structures \cite{10.1145/1831708.1831717}. SCMs model causal mechanisms through structural assignments and support well-defined interventions \cite{10.1145/3635709}, while DAG-based representations make causal assumptions explicit. In some approaches, program dependence relations are treated as approximations of causal precedence, allowing PDGs to serve as structural proxies for causal reasoning \cite{lee2021causalprogramdependenceanalysis}. These representations establish the structural assumptions required for subsequent identification and estimation of causal effects.

\textbf{\textit{2. Causal Structure Discovery Layer.}}
The causal structure discovery layer addresses how causal graphs are inferred from data when the underlying dependency structure is not explicitly specified. Representative approaches include constraint-based discovery \cite{10.1109/ICSE-NIER58687.2023.00018}, score-based discovery \cite{10556461}, and intervention-enhanced discovery methods \cite{10.1109/ASE56229.2023.00106}. These techniques estimate DAG structures from observational and/or interventional data by exploiting statistical independence tests, optimization criteria, or prior knowledge constraints. They provide data-driven approximations of causal structures that can subsequently support identification and estimation of causal effects.

\textbf{\textit{3. Causal Identification Layer.}}
The causal identification layer determines whether and under what conditions causal effects are identifiable from a given causal structure. Its primary goal is to establish valid estimands and specify the variables that must be controlled to eliminate confounding.
Representative mechanisms include graphical identifiability analysis based on the backdoor criterion~\cite{6569724}, d-separation~\cite{kucuk2021improvingfaultlocalizationintegrating}, and adjustment set selection~\cite{10.1145/3607184}. These approaches analyze the causal graph to determine whether a target causal effect is identifiable and to derive sufficient adjustment sets that block spurious backdoor paths.

In addition, some studies operationalize explicit interventions, such as do($X{=}x$) interventions~\cite{10.1109/ICSE-NIER58687.2023.00018}, input perturbation~\cite{10.1145/3377811.3380377}, fault injection~\cite{10.1145/3318464.3389694}, and mutation-based interventions~\cite{oh2021effectivelysamplinghigherorder}, to approximate hypothetical manipulations in software systems. These intervention realizations support identification by enabling controlled manipulation of treatment variables consistent with the underlying causal assumptions.

Together, graphical identification and intervention operationalization provide a principled basis for establishing identifiable causal queries prior to effect estimation.

\textbf{\textit{4. Causal Effect Estimation Layer.}}
The causal effect estimation layer focuses on computing the magnitude of identifiable causal effects once valid estimands have been established. This layer concerns statistical procedures and computational techniques used to quantify causal relationships under the assumed causal model.

Representative estimation techniques include regression adjustment \cite{6569724}, Double Machine Learning (DML) \cite{10.1109/ASE56229.2023.00106}, Instrumental Variable (IV) estimation \cite{foster2025usingcausalinferencetest}, simulation-based approaches \cite{10.1109/ICSE-NIER58687.2023.00018}, and counterfactual-based estimation \cite{Li2025EnhancingFL}. These techniques operationalize the estimand derived in the identification layer and provide numerical effect estimates under specific modeling assumptions.

Target causal estimands commonly reported in the reviewed studies include the Average Treatment Effect (ATE), Conditional Average Treatment Effect (CATE), Average Causal Effect (ACE), and Direct Causal Effect (DCE). The choice of estimand determines whether the analysis focuses on population-level effects, subgroup heterogeneity, or path-specific causal influence.

Together, these estimation mechanisms enable quantitative evaluation of causal impact in software testing and debugging scenarios.

\subsubsection{\textbf{Underlying Assumptions of Causal Inference Techniques}}

Across the reviewed studies, the application of causal inference in software testing relies on recurring methodological assumptions aligned with the SCM-based pipeline: causal representation, structure discovery, identification (including intervention), and effect estimation.

\textbf{\textit{1. Assumptions for Causal Representation.}}

\begin{itemize}
    \item \textbf{Causal model adequacy.} The specified or learned causal model (e.g., SCM or DAG) sufficiently captures the relevant variables and causal relationships of the system under test \cite{10132179,10.1109/ICSE-NIER58687.2023.00018,10.1145/3607184}.
    
    \item \textbf{Structural stability.} The underlying causal mechanisms remain invariant across executions, inputs, and admissible interventions \cite{10132179,10.1145/3318464.3389694}.
    
    \item \textbf{Acyclicity (for DAG-based models).} The causal structure forms a directed acyclic graph, ensuring well-defined effect propagation \cite{monjezi2023informationtheoretictestingdebuggingfairness,lee2021causalprogramdependenceanalysis,6569724}.
\end{itemize}

\textbf{\textit{2. Assumptions for Causal Structure Discovery.}}

\begin{itemize}
    \item \textbf{Faithfulness.} Statistical dependencies observed in execution data faithfully reflect the underlying causal structure, enabling consistent graph recovery \cite{10.1109/ASE56229.2023.00106,oh2021effectivelysamplinghigherorder}.
    
    \item \textbf{Discovery validity.} Available data and the discovery algorithm provide sufficient signal, under their assumptions, to recover the intended causal structure \cite{lee2021causalprogramdependenceanalysis}.
\end{itemize}

\textbf{\textit{3. Assumptions for Identification and Intervention.}}

\begin{itemize}
    \item \textbf{Intervention validity.} Testing interventions correspond to semantically meaningful \emph{do}-manipulations and do not introduce unintended side effects \cite{10.1145/3607184,10.1109/ICSE-NIER58687.2023.00018,monjezi2023informationtheoretictestingdebuggingfairness,10.1145/3635709}.
    
    \item \textbf{No unmeasured confounding.} All relevant common causes of treatment and outcome are measured or properly controlled \cite{6569724,10.1145/1831708.1831717}.
    
    \item \textbf{Effect identifiability.} The causal effect of interest is identifiable from the assumed structure and available data (e.g., via valid adjustment sets) \cite{10.1109/ICSE-NIER58687.2023.00018,10.1109/ASE56229.2023.00106,10.1145/3635709}.
    
    \item \textbf{Treatment well-definedness.} The treatment corresponds to a clearly specified and consistently implemented test configuration \cite{Feyzi_2019,6569724}.
\end{itemize}

\textbf{\textit{4. Assumptions for Effect Estimation.}}

\begin{itemize}
    \item \textbf{Positivity (overlap).} Comparable executions exist for each treatment level, enabling reliable effect estimation \cite{kucuk2021improvingfaultlocalizationintegrating,Feyzi_2019}.
    
    \item \textbf{Estimator adequacy.} The chosen estimation method and underlying models are correctly specified and sufficiently accurate to yield consistent estimates \cite{kucuk2021improvingfaultlocalizationintegrating,Feyzi_2019}.
    
    \item \textbf{Data stability.} The data-generating process remains stable across executions or reuse scenarios, avoiding distortion from distributional shifts or mechanism changes \cite{lee2021causalprogramdependenceanalysis,10.1145/3696630.3731613}.
\end{itemize}
\begin{table*}[t]
\centering
\small
\caption{Causal Inference Techniques and Their Roles in Software Testing (RQ2)}
\begin{tabularx}{\textwidth}{|
>{\hsize=0.4\hsize}X|
>{\hsize=1.1\hsize}X|
>{\hsize=0.99\hsize}X|
>{\hsize=1.5\hsize}X|}
\hline
\textbf{Layer} & \textbf{Elements and Methods/Tools} & \textbf{Representative Studies} & \textbf{Role} \\ \hline

Causal representation layer
&
\textbf{Causal Model Formalism}

SCM, 
DAG,
PDG
&
\cite{10.1145/1831708.1831717}
\cite{10.1145/2025113.2025136}
\cite{7102597}
\cite{10.1145/3607184}
\cite{10132179}
\cite{10.1145/3696630.3731613}
\cite{11260638}
\cite{10.1145/3318464.3389694}
\cite{Feyzi_2019}
\cite{foster2025usingcausalinferencetest}
\cite{10638595}
\cite{10556461}
\cite{10.1145/3635709}
\cite{10.1109/ICSE-NIER58687.2023.00018}
\cite{10.1109/ASE56229.2023.00106}
\cite{kucuk2021improvingfaultlocalizationintegrating}
\cite{lee2021causalprogramdependenceanalysis}
\cite{Li2025EnhancingFL}
\cite{monjezi2023informationtheoretictestingdebuggingfairness}
\cite{oh2021effectivelysamplinghigherorder}
\cite{9240635}
\cite{Poskitt_2023}
\cite{6569724}
&
Encodes test-relevant program entities and dependencies as an explicit causal structure, making causal assumptions and intervention targets clear for downstream identification and estimation.
\\ \hline

Causal structure discovery layer
&
\textbf{Automatic Causal Graph Discovery}

Causal structure discovery

Constraint-based discovery

Score-based discovery

Intervention-driven discovery

Hybrid discovery
&
\cite{11260638}
\cite{10.1145/3318464.3389694}
\cite{10556461}
\cite{10.1145/3635709}
\cite{10.1109/ICSE-NIER58687.2023.00018}
\cite{10.1109/ASE56229.2023.00106}
\cite{lee2021causalprogramdependenceanalysis}
\cite{oh2021effectivelysamplinghigherorder}
&
Learns the causal graph from execution paths and/or intervention data when no a priori model is available, providing a data-driven structure to support following identifiability analysis and effect estimation.
\\ \hline

Causal identification layer
&
\textbf{Graphical Identifiability Analysis}

Backdoor criterion

d-separation

Adjustment set selection

Causal path analysis
&
\cite{10.1145/1831708.1831717}
\cite{10.1145/2025113.2025136}
\cite{7102597}
\cite{10.1145/3607184}
\cite{10.1145/3696630.3731613}
\cite{10.1145/3318464.3389694}
\cite{Feyzi_2019}
\cite{foster2025usingcausalinferencetest}
\cite{10638595}
\cite{10.1145/3635709}
\cite{6227169}
\cite{kucuk2021improvingfaultlocalizationintegrating}
\cite{lee2021causalprogramdependenceanalysis}
\cite{Li2025EnhancingFL}
\cite{9240635}
\cite{Poskitt_2023}
\cite{6569724}
&
Determines whether a testing effect is identifiable under the graph and derives sufficient adjustment sets to control confounding, thereby defining a valid estimand for effect estimation.
\\ 
\cline{2-4}

&
\textbf{Intervention Realization}

do(X=x) intervention

Input perturbation

Fault injection

Mutation-based intervention

Chaos engineering intervention
&
\cite{10.1145/1831708.1831717}
\cite{10.1145/3607184}
\cite{10132179}
\cite{11260638}
\cite{10.1145/3318464.3389694}
\cite{Feyzi_2019}
\cite{foster2025usingcausalinferencetest}
\cite{10556461}
\cite{10.1145/3635709}
\cite{10.1109/ICSE-NIER58687.2023.00018}
\cite{10.1109/ASE56229.2023.00106}
\cite{10.1145/3377811.3380377}
\cite{lee2021causalprogramdependenceanalysis}
\cite{Li2025EnhancingFL}
\cite{10.1145/3663529.3663807}
\cite{monjezi2023informationtheoretictestingdebuggingfairness}
\cite{oh2021effectivelysamplinghigherorder}
\cite{Poskitt_2023}
\cite{6606573}
&
Conduct causal manipulations in the SUT (e.g., via perturbation, injection, mutation) to approximate do-interventions, enabling controlled comparisons that support causal effect identification in debugging and testing.
\\ \cline{2-4}
\hline

Causal effect estimation layer
&
\textbf{Estimation Techniques}

Regression adjustment

Double Machine Learning (DML)

Instrumental Variable estimation

Simulation-based approach

Counterfactual effect estimation

&
\cite{10.1145/1831708.1831717}
\cite{10.1145/2025113.2025136}
\cite{7102597}
\cite{10.1145/3607184}
\cite{10.1145/3696630.3731613}
\cite{11260638}
\cite{Feyzi_2019}
\cite{foster2025usingcausalinferencetest}
\cite{10638595}
\cite{10556461}
\cite{10.1145/3635709}
\cite{10.1109/ICSE-NIER58687.2023.00018}
\cite{6227169}
\cite{10.1109/ASE56229.2023.00106}
\cite{kucuk2021improvingfaultlocalizationintegrating}
\cite{lee2021causalprogramdependenceanalysis}
\cite{Li2025EnhancingFL}
\cite{monjezi2023informationtheoretictestingdebuggingfairness}
\cite{oh2021effectivelysamplinghigherorder}
\cite{9240635}
\cite{6569724}
&
Quantifies the identified causal effect using statistical/ML estimators, measures the causal effect size of a testing intervention on the outcome.
\\ \cline{2-4}

&
\textbf{Target Causal Estimands}

ATE,
CATE,
ACE,
DCE
&
\cite{10.1145/1831708.1831717}
\cite{7102597}
\cite{10.1145/3607184}
\cite{10.1145/3696630.3731613}
\cite{11260638}
\cite{foster2025usingcausalinferencetest}
\cite{10638595}
\cite{10.1145/3635709}
\cite{10.1109/ICSE-NIER58687.2023.00018}
\cite{6227169}
\cite{10.1109/ASE56229.2023.00106}
\cite{kucuk2021improvingfaultlocalizationintegrating}
\cite{monjezi2023informationtheoretictestingdebuggingfairness}
\cite{9240635}
\cite{6569724}
&
Specifies the causal effect, aligning the identification assumptions with the estimation procedure and clarifying how results should be interpreted for testing.
\\ \cline{2-4}
\hline
\end{tabularx}
\end{table*}

Table~\ref{tab:layer_phase_cross} shows the distribution of causal inference layers across testing phases.

\noindent
\begin{minipage}{0.5\textwidth}
\centering
\captionof{table}{Cross-distribution of Causal Layers and Test Phases}
\label{tab:layer_phase_cross}
\footnotesize
\setlength{\tabcolsep}{4pt}

\begin{tabular}{|p{0.3\linewidth}|c|c|c|c|}
\hline
\textbf{Layer\textbackslash Test phases} & 
\textbf{Design} & 
\textbf{Execution} & 
\textbf{Interpretation} & 
\textbf{Debugging} \\
\hline
Causal representation & 8 & 2 & 15 & 13 \\
\hline
Causal structure discovery & 4 & 2 & 3 & 4 \\
\hline
Causal identification & 8 & 2 & 17 & 17 \\
\hline
Causal effect estimation  & 6 & 2 & 15 & 13 \\
\hline
\end{tabular}
\end{minipage}
\subsection{RQ3:  What challenges limit the application of causal inference in software testing?}

The reviewed studies reveal recurring challenges across the causal inference pipeline, spanning representation, discovery, identification/intervention, and estimation. These risks collectively constrain the practical reliability and scalability of causal inference techniques in software testing.

\textbf{\textit{Representation-Level Risk: Model Dependence and Abstraction Gap.}}
Most approaches rely heavily on correctly specified causal models. In practice, translating program semantics, hidden states, program interactions, and runtime configurations into causal variables and edges requires abstraction and simplification. Program dependence does not always coincide with true causal influence, and omitted variables or interactions can lead to misspecified graphs. Such representation errors propagate to downstream identification and estimation, undermining the validity of causal conclusions~\cite{10132179,7102597,10.1109/ICSE-NIER58687.2023.00018}.

\textbf{\textit{Discovery-Level Risk: Structural Uncertainty and Scalability.}}
When causal structure is automatically learned, discovery procedures depend on strong assumptions (e.g., causal sufficiency and faithfulness) and adequate data diversity. Sparse failures, cyclic dependencies, and limited intervention diversity can result in unstable or inconsistent learned structures. Furthermore, the combinatorial search space over possible graphs and sensitivity to tool configurations introduce substantial computational overhead, limiting scalability to larger systems \cite{10.1145/3635709}.

\textbf{\textit{Identification-Level Risk: Unverifiable Assumptions and Intervention Constraints.}}
Causal identification requires that confounding be sufficiently controlled. In software systems, hidden states, partially observable executions, and correlated behaviors make these assumptions difficult to verify. Many approaches depend on explicit interventions (e.g., input perturbation, mutation, or fault injection), which may be infeasible, unsafe, or inconsistent with system constraints. Counterfactual constructions derived from such interventions may also be unrealistic, further limiting the credibility of identified causal relationships \cite{10.1145/3318464.3389694,10.1109/ICSE-NIER58687.2023.00018,Poskitt_2023}.

\textbf{\textit{Estimation-Level Risk: Limited Support and Model Sensitivity.}}
Even when effects are theoretically identifiable, estimation remains fragile under limited support. Software testing contexts often exhibit sparse failure cases and imbalanced data coverage, leading to high-variance or biased effect estimates. In addition, estimation outcomes depend on modelling choices, making causal effect metrics (e.g., ATE or CATE) sensitive to data quality \cite{10.1145/3696630.3731613,10.1145/3635709}.

Across all layers, limited generalisability remains a concern. Many studies rely on small benchmarks or simulated environments, leaving their applicability to large real-world systems uncertain.

\subsection{RQ4: What are the research gaps and future directions?}

Future research directions can focus on improving representation, discovery, estimation, and practical adoption in software testing.

\textbf{\textit{More Accurate and Automated Causal Representation.}}
Future work should explore systematic and automated ways to build causal models of software systems. This includes clearer mappings from program structure to causal variables and edges. Combining static analysis, dynamic analysis, and data-driven methods may help reduce modeling errors and improve downstream reliability.

\textbf{\textit{Scalable Causal Structure Discovery for Large Systems.}}
Improving the scalability of causal structure discovery is an important direction. Future methods should better handle limited failure cases and noisy data. Integrating domain knowledge with data-driven learning may help reduce complexity and improve stability in large software systems.

\textbf{\textit{Practical and Safe Intervention Mechanisms.}}
Future research should design intervention strategies that are feasible and safe in real software environments. This includes better intervention selection, input perturbations, and realistic counterfactual test generation. Making interventions more practical can improve the applicability of causal testing techniques.

\textbf{\textit{Robust Effect Estimation under Real-World Constraints.}}
Causal effect estimation should be made more robust under incomplete data, hidden confounders, and unstable program traces. Future work may focus on uncertainty detection, bias analysis, and adaptive data collection.

\textbf{\textit{Stronger Validation and Better Integration into Practice.}}
Future studies should improve validation and evaluation practices, including broader benchmarks, clearer uncertainty report, and systematic comparison across systems. In addition, better integration with debugging workflows and practical tool chains will be important for better adoption in real-world software engineering.

\section{Threats to Validity}
\textbf{\textit{Search and Selection Bias.}}
Our search strategy relied on causality- and testing-related keywords and major digital libraries. Relevant studies may have been missed if they did not explicitly use these terms in titles or abstracts. To mitigate these risks, we complemented database searches with backward and forward snowballing. Nonetheless, some relevant work may remain uncaptured.

\textbf{\textit{Construct Validity and Classification Bias.}}
Data extraction, layer mapping, and taxonomy construction required interpretive judgment, particularly when assigning studies to stages of the causal inference pipeline. This introduces the risk of subjective interpretation and misclassification. We used a structured extraction form and predefined quality criteria to improve consistency, but complete objectivity cannot be guaranteed.

\textbf{\textit{Evidence Strength Variability.}}
Quality assessment was used to contextualize evidence rather than exclude studies. As a result, the included set may contain variability in methodological rigor, which could influence the strength of synthesized conclusions.


\section{Conclusion}



This paper presents that causal inference is primarily applied to result interpretation and debugging, with growing use of intervention-based identification and effect estimation. However, recurring issues, including model misspecification, strong identification assumptions, scalability limitations, and limited real-world validation, constrain practical impact. By organizing existing work through a layered causal inference perspective, this review provides a structured foundation for advancing more robust, scalable, and empirically grounded causal testing methodologies.

\bibliographystyle{abbrv}
\bibliography{sigproc}  
\end{document}